\documentclass[aps,prl,twocolumn,superscriptaddress,showpacs]{revtex4-1}
\usepackage{graphicx}
\usepackage{dcolumn}
\usepackage{bm}
\usepackage{xcolor}
\usepackage{multirow}
\usepackage{subfigure}
\usepackage{hyperref}
\usepackage{amssymb}
\usepackage{color}
\usepackage{graphics}\usepackage{epsfig}\usepackage{subfigure}
\usepackage{footnote}
\usepackage{amsmath}

\begin{document}
\title{Chiral spin texture in the charge-density-wave phase of the correlated metallic Pb/Si(111) monolayer}
\author{C.~Tresca}
\affiliation{Department of Physical and Chemical Sciences and SPIN-CNR, University of L'Aquila, Via Vetoio 10, I-67100 L'Aquila, Italy}
\affiliation{Institut des Nanosciences de Paris, Sorbonne Universit\'es-UPMC univ Paris 6 and CNRS-UMR 7588, 4 place Jussieu, F-75252 Paris, France}
\author{C.~Brun}
\email{christophe.brun@upmc.fr}
\author{T.~Bilgeri}
\author{G.~Menard}
\author{V.~Cherkez}
\author{R.~Federicci}
\author{D.~Longo}
\author{F.~Debontridder}
\author{M.~D'angelo}
\affiliation{Institut des Nanosciences de Paris, Sorbonne Universit\'es-UPMC univ Paris 6 and CNRS-UMR 7588, 4 place Jussieu, F-75252 Paris, France}
\author{D.~Roditchev}
\affiliation{Institut des Nanosciences de Paris, Sorbonne Universit\'es-UPMC univ Paris 6 and CNRS-UMR 7588, 4 place Jussieu, F-75252 Paris, France}
\affiliation{Laboratoire de physique et d'\'etude des mat\'eriaux, LPEM-UMR8213/CNRS-ESPCI ParisTech-UPMC,
10 rue Vauquelin, F-75005 Paris, France}
\author{G.~Profeta}
\affiliation{Department of Physical and Chemical Sciences and SPIN-CNR, University of L'Aquila, Via Vetoio 10, I-67100 L'Aquila (Italy)}
\author{M.~Calandra}
\email{matteo.calandra@insp.upmc.fr}
\affiliation{Institut des Nanosciences de Paris, Sorbonne Universit\'es-UPMC univ Paris 6 and CNRS-UMR 7588, 4 place Jussieu, F-75252 Paris, France}
\author{T.~Cren}
\affiliation{Institut des Nanosciences de Paris, Sorbonne Universit\'es-UPMC univ Paris 6 and CNRS-UMR 7588, 4
place Jussieu, F-75252 Paris, France}


\begin{abstract}
We investigate the 1/3 monolayer $\alpha$-Pb/Si(111) surface by scanning tunneling spectroscopy (STS)  and fully relativistic first-principles calculations. We study both the high-temperature $\sqrt{3}\times\sqrt{3}$ and low-temperature $3\times 3$ reconstructions and show that, in both phases, the spin-orbit interaction
leads to an energy splitting as large as $25\%$ of the valence-band bandwidth. Relativistic effects, electronic correlations and  Pb-substrate interaction cooperate to stabilize a correlated low-temperature paramagnetic phase with well-developed lower and upper Hubbard bands coexisting with $3\times3$ periodicity. By comparing the Fourier transform of STS conductance maps at the Fermi level with calculated quasiparticle interference from non-magnetic impurities, we demonstrate the occurrence of two large hexagonal Fermi sheets with in-plane spin polarizations and opposite helicities.
\end{abstract}.

\maketitle

Two-dimensional materials with spin-polarized surface states are promising candidates for spin-charge current conversion via the Edelstein effect \cite{FertEmergent}. 
Furthermore, efficient long-range spin coherent applications \cite{Wolf1488} require
 spin-transport protected from backscattering. A first step in this direction using strong Rashba spin-orbit coupling (SOC) has been made in some heavy-group atoms grown epitaxially on group IV surfaces, like $\beta$-Pb/Ge(111) \cite{10.1038/ncomms1016, 2016Ren} or Au/Si(111) \cite{10.1038/srep01826}. Surprisingly, in less dense 1/3 monolayer (ML) phases, effects produced by SOC were overlooked while reversible phase transitions as a function of temperature were found \cite{Carpinelli1996, Carpinelli1997, PhysRevB.69.241307, Cudazzo2008747}, accompanied by metal-insulator transitions \cite{Cortes,Modesti2007,PhysRevLett.98.086401} and possible magnetic orderings \cite{ncomms2617, PhysRevLett.111.106403}.


\begin{figure*}
\centering 
{\includegraphics[width=\textwidth]{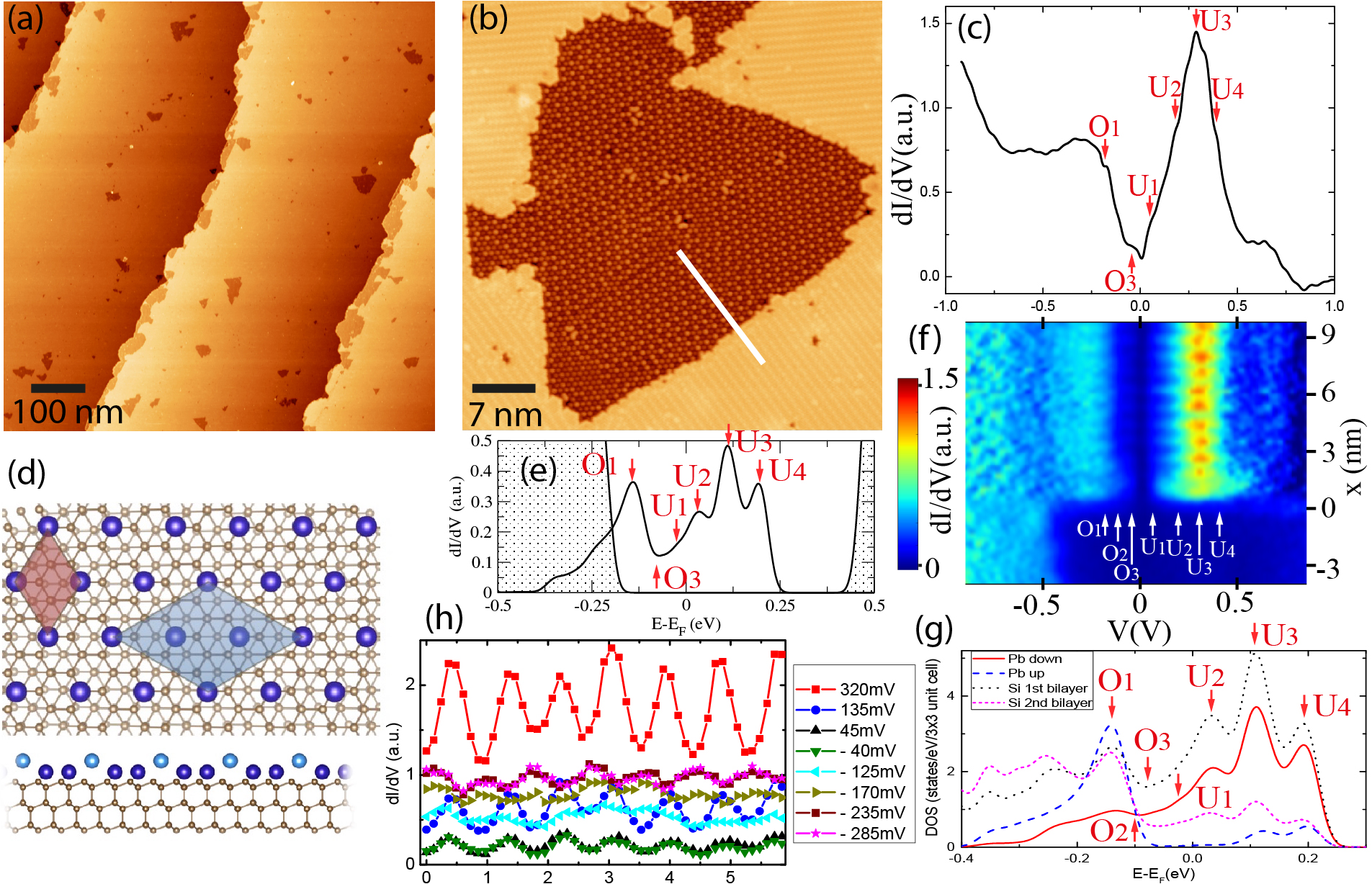}}
\caption{Measurements at $T=4.2$~K. (a) $900\times 900$ nm$^2$ STM topographic map of the sample : small 3$\times$3 areas surrounded by large $\sqrt{7}\times\sqrt{3}$ domains. Scanning parameters $V=-1.0$~V, I=20~pA. (b) $48\times 48$ nm$^2$ topography of a 3$\times$3 area showing Pb up and down atoms. (c) $dI/dV(V)$ conductance spectra, measured by STS between $[-0.9 ,+0.9]$~V inside a 3$\times 3$ domain in a zone free-of-defects, spatially averaged over 9~nm$^2$. Set-point: $V=-1.0$~V, I=200~pA. (d) Top: top-view of the high-temperature $\sqrt3 \times \sqrt3 ~R30^{\circ}$ (red) and low-temperature $3\times3$ (blue) unit cells of 1/3~ML Pb/Si(111). Bottom: Side-view of the $3\times3$ $\alpha$-phase. Small (large) circles represent Si (Pb) atoms. (e) Computed STS spectrum for the 3$\times$3 $\alpha$-phase using Tersoff-Hamann (TH) theory \cite{TersoffHamann} in the DFT+U approximation, including SOC effects for a supercell having three Si double-layers. The grey region is the TH calculation for a semi-infinite Si(111) slab. (f) Color plot representing the conductance of 62 $dI/dV(V)$ spectra measured along the 13.8~nm white line shown in panel (b), across the interface between a 3$\times$3 and a $\sqrt{7}\times\sqrt{3}$ domain. The right axis indicates the distance $x$ from the boundary (being at x=0 nm) at which each spectrum is measured. The voltage axis is identical to the one used in (c) enabling a direct comparison of all spectral features. Each spectrum is the average of 15 spectra measured over a width of 5~nm perpendicular to the white line. The 3$\times$3 bandstructure establishes sharply over less than 1~nm from the interface. Periodic oscillations appear along $x$ in the range $[-0.3,+0.3]$~V. They reflect the different LDOS at the Pb-up and down atoms, seen in panel (g) presenting the calculated LDOS projected on atomic sites. Panel (h) presents cuts at various energies in map (f) enabling to better see the oscillations in (f).}\label{STM}
\end{figure*}


At room temperature 1/3 ML coverage of Pb or Sn grown on top of Si(111) or Ge(111) display an isoelectronic $\sqrt{3}\times\sqrt{3}~R30^{\circ}$ reconstruction. The host atoms occupy the T$_4$ sites atop the substrate surface in an hexagonal array \cite{Carpinelli1996}, the so-called $\alpha$-phase (see Fig.~\ref{STM}d). 
The three Si dangling bonds in the top Si layer next to the metal atom are saturated and a free unsaturated electron is left at each T4 site.
Thus these systems are expected to present a half-filled surface band well separated from the Si bands. 
This expectation is however in contrast with experimental evidences of an insulating ground state in Sn on Ge(111) \cite{Cortes} and Si(111) \cite{Modesti2007,Odobescu2017}. 

The reason of this behaviour has been attributed to electronic correlations \cite{PhysRevLett.98.086401,Hansmann2013b}. 
At low temperature the situation is complicated by the presence in most systems of 
a reversible structural transition from $\sqrt 3 \times \sqrt 3$ to $3 \times 3$ periodicity
\cite{Carpinelli1996, Carpinelli1997,Brihuega2005,Brihuega2007}. 
Although its origin has been extensively debated, it is still not clear yet whether this transition is due to the freezing of an out-of-plane phonon mode \cite{Cudazzo2008747} or produced by long-range electronic correlations \cite{Hansmann2013a,Hansmann2013b}, both favouring a charge-density-wave (CDW) having the so-called 1Up-2Down atoms configuration. Surprisingly, despite the presence of heavy atoms and substrate-induced broken inversion symmetry, only one theoretical work  (restricted to the $\sqrt{3}\times \sqrt{3}$ reconstruction) 
treats {\it ab initio} and on equal footing relativistic and correlation effects as well as the interaction with the substrate\cite{Badrtdinov2016}. Finally, despite many theoretical calculations for 1/3 ML Pb/Si(111) \cite{Hansmann2013b,Cudazzo2008747,Badrtdinov2016}, the small size of the $3 \times 3$
domains has prevented, up to now, the experimental spectroscopy of their electronic structure.

In this work we study the 1/3 ML $\alpha$-Pb/Si(111) by scanning tunneling microscopy/spectroscopy (STM/STS) and by fully-relativistic Density Functional Theory (DFT). Experimental results down to 300~mK reveal the $3 \times 3$ groundstate to be a correlated metal not undergoing a Mott transition. It is characterized by a highly depressed density of states (DOS) at the Fermi level $E_F$ where well-defined quasiparticles exist, resulting in two large 
Fermi surfaces with dominant in-plane spin polarization and opposite helicities.


The preparation of the $\alpha$-Pb/Si(111) surface is described in supplemental material (SM) \cite{supplemental}. It is well-known that it is not possible to grow very large $\sqrt3 \times \sqrt3$ areas with a low-density of defects \cite{Brihuega2005,Brihuega2007} in contrast to Pb/Ge(111), Sn/Ge(111) or Sn/Si(111). Below 86~K the $\sqrt3 \times \sqrt3$ regions transit to a $3 \times 3$ structure \cite{Brihuega2005,Brihuega2007} (see the atomic positions in Fig.~\ref{STM}d). The samples were transfered under ultra-high vacuum to the cold STM head where measurements were performed at 0.3~K, 2~K and 4~K with metallic PtIr or W tips. The $dI/dV(V)$ spectra shown hereafter were obtained by numerical derivation of the raw $I(V)$ curves. Negative (positive) bias voltage corresponds to occupied (empty) sample states.

A $900 \times 900$~nm$^2$ STM topography of the studied samples, measured at 4~K, is shown in Fig.~\ref{STM}a. $3\times 3$ patches of various sizes are surrounded by metallic $\sqrt{7}\times\sqrt{3}$ domains. The latter ones enable a good electrical connectivity and efficient charge evacuation \cite{Brun2017}. $3\times 3$ patches of size 25-100~nm, such as the main one seen in Fig.~\ref{STM}b, were chosen for the spectroscopic studies. This size range is similar to the one presented in \cite{Brihuega2005}. Intrinsic $3\times 3$ spectroscopic features were measured far enough from boundaries with neighboring $\sqrt7 \times \sqrt3$ domains.

A characteristic $dI/dV$ spectrum measured locally in clean 3$\times$3 areas is shown in Fig.~\ref{STM}c. It reveals the existence of a strongly depressed DOS at $E_F$ not reaching zero demonstrating the correlated semimetallic character of the surface. The most prominent feature is a peak observed at +290~meV labelled $U_3$. Robust kinks were always observed at -170~meV and -60~meV labelled $O_1$ and $O_3$. At larger binding energy ($-420<V<-250$~meV) the conductance increases forming a broad peak with internal fine features. Then the DOS remains flat with small oscillations until $V\simeq-660$~meV and increases strongly for larger energy ($V<-660$~meV) due to Si bands \cite{PhysRevB.80.125326}. In the unoccupied states we also observe several reproducible small kinks around the peak $U3$ at +55~meV, +180~meV and +400~meV denoted respectively $U1$, $U2$ and $U4$. It is followed at larger energy ($V>400$~meV) by a continuous drop in conductance until about 550~meV. The conductance remains low but non-zero until 650~meV, where it starts to drop, leading to negative $dI/dV$ above $V\simeq 750$~meV. The conductance remains negative below 1~V and strongly increases above 1~V further signaling Si bands (see II in SM for additionnal data and discussion of the small features \cite{supplemental}).

In principle, the absolute energies of these features must be taken with caution as the electric field of the STM tip is known to induce a band bending effect at the surfaces of gapped solids like semiconductors \cite{PhysRevB.80.125326} or Mott systems \cite{Battisti}. However, here less than $10\%$ of the surface is covered by 3$\times$3 areas, while the remaining is the metallic $\sqrt{7}\times\sqrt{3}$ phase becoming superconducting below $1.5~K$ \cite{Brun2017,BrunNatphys}. Furthermore, as 3$\times$3 areas are found to be semimetallic and not insulating, small band bending effect is expected. The comparison between STS, ARPES and DFT+U in 1/3ML Sn/Si(111) shows negligible band bending for the occupied states and a small effect for the unoccupied states \cite{Modesti2007}. The most important contributions that we found to influence the LDOS of 3$\times$3 areas are (i) local or extended defects and (ii) dynamical motion of Pb atoms below the tip (see section II in \cite{supplemental}).

\begin{figure}[ht]
\centering 
\includegraphics[width=0.95\linewidth]{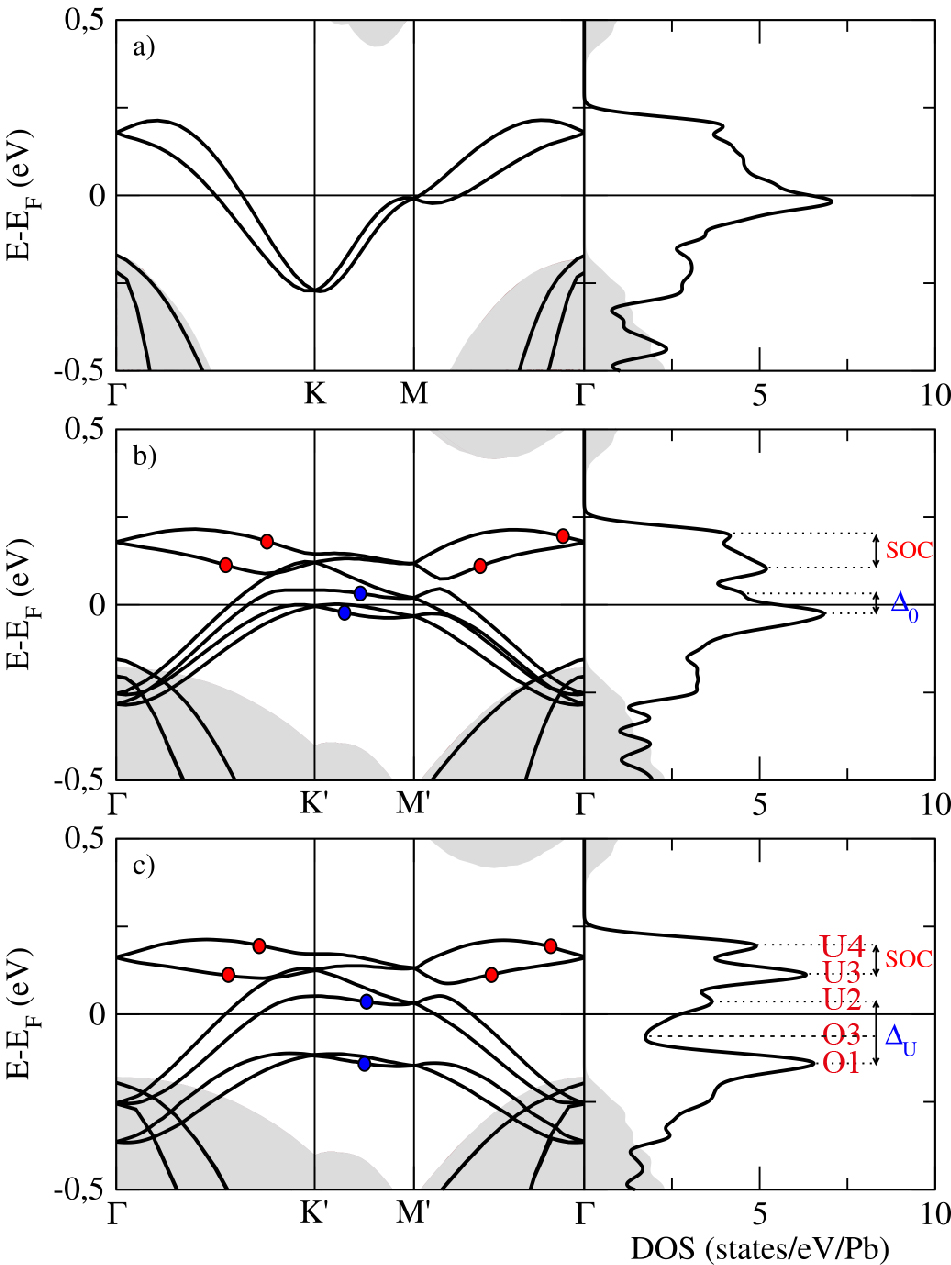}
\caption{Fully relativistic electronic structure for the $\sqrt{3}\times\sqrt{3} {\rm R}(30^\circ)$ (a) and  $3\times3$ (b) phases with $U=0$ eV. In (c)  $3\times3$ with $U=1.75$ eV. 
Dots in (b) and (c) label the {\bf k}-points leading to peaks in the DOS (right panels). The projected Si bulk band structure (and DOS) is indicated by gray areas.}\label{r3xr3_band_SO}.
\end{figure}

To rationalize the experimental findings we performed theoretical calculations using \textsc{Quantum-Espresso}~\cite{QEcode}. We included relativistic effects, non-collinear magnetism in the DFT+U approximation and the Si(111) substrate (see SM \cite{supplemental} and Ref. \cite{Cudazzo2008747} for more details).   We first consider 1/3 ML Pb on Si(111) in the $\sqrt{3}\times\sqrt{3}~R30^{\circ}$ structure within the Spin Polarized Generalized Gradient Approximation (GGA). 
While the structural properties are not affected  by relativistic effects (see SM \cite{supplemental} and Ref. \cite{Cudazzo2008747}), 
SOC has a strong impact on the electronic structure inducing a band splitting that can be as large as  $1/4$ of the bandwidth as shown in Fig.~\ref{r3xr3_band_SO}a.
The electron-electron interaction treated in the DFT+U approximation or using the HSE06 has minor effects on the $\sqrt{3}\times\sqrt{3}$ surface-band electronic structure\cite{supplemental}. 

To understand the microscopic mechanisms behind the 3$\times$3 reconstruction, 
we first neglect SOC and study the lattice deformation within the GGA and LDA approximation. Using GGA, we find a stabilization of the $3\times3$ reconstruction (energy gain 5~meV per Pb atom, 
difference in height of Pb atoms $\Delta h\sim$0.24\AA ) in agreement with previous results \cite{Cudazzo2008747}. 
Despite the indistinguishable electronic structure, LDA does not lead to a structural transition, ruling out Fermi surface effects. 
Although the structural 3$\times$3 distortion is well reproduced in paramagnetic GGA neglecting SOC, the resulting surface DOS is clearly metallic with no depressed DOS at $E_F$ (see SM \cite{supplemental}), in qualitative disagreement with experiment. Including SOC splits the topmost peak in the empty DOS by $\approx 0.09$ eV, generating a three-peaks structure, but still leads to a metallic state (see Fig.~\ref{r3xr3_band_SO}b).
Within GGA, we also considered possible magnetic instabilities. Even if we could stabilize different magnetic solutions, those were always higher in energy than the paramagnetic one \cite{supplemental} .
\begin{figure*}[ht]
\centering 
\includegraphics[width=0.95\textwidth]{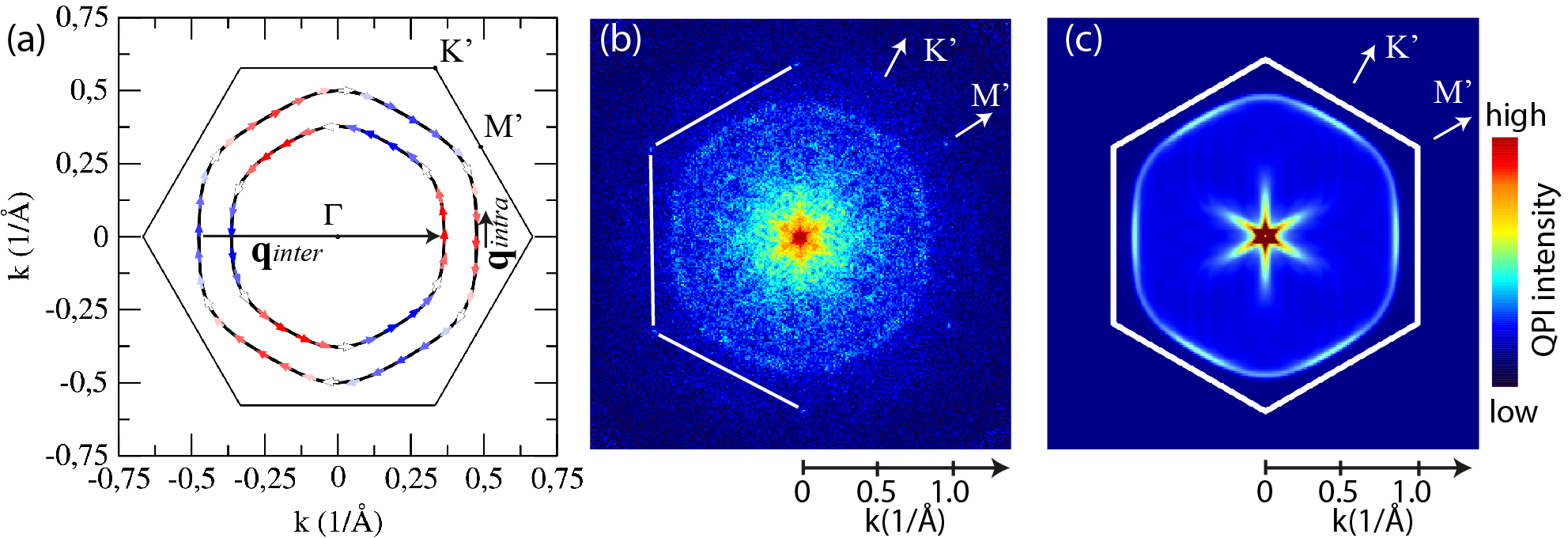}
\caption{(a) DFT+U Fermi surface of Pb-$3\times3$/Si(111) including spins polarization (arrows). The out-of-plane spin component is at most 1$\%$ of the in-plane one. White arrows: 100$\%$ in-plane polarization, blue and red arrows: opposite out-of-plane components. Black arrows $\bold{q}_{inter}$ and $\bold{q}_{intra}$ : scattering vectors corresponding to scalar impurities. (b) Symmetrized Fourier transform of a $60\times60$~nm$^2$ $dI/dV(V=0)$ map measured by STS at $T=0.3$~K and $B=0.5$~T, corresponding to quasiparticle interference at $E=E_F$. (c) Calculated quasiparticle interference map at $E=E_F$ assuming pure scalar impurity scattering.}\label{Fig:FTSTS}. 
\end{figure*}
We then consider the relativistic rotational-invariant formulation of 
the GGA+U approximation \cite{PhysRevB.71.035105}. We evaluated the $U$ parameter self-consistently and obtained $U\sim 1.75$ eV. We optimized the internal parameters and again found that the paramagnetic 3$\times$3 phase is the ground state (8~meV/Pb energy gain and $\Delta h \sim 0.32~\AA $). As the energy gain is larger at $U\sim 1.75$ eV than at $U= 0$ eV, it means that the 
local Coulomb repulsion enhances the distortion.
Fig.~\ref{r3xr3_band_SO}c shows the electronic structure. The Hubbard repulsion lowers the energies of half of the occupied bands reducing the DOS near $E_F$. 
It is possible to probe directly the SOC ($\Delta_{\rm SOC}$) and Hubbard ($\Delta_{\rm U}$) splittings at selected {\bf k}-points by looking at the corresponding features induced in the empty and occupied DOS. As presented in the right panel of fig.~\ref{r3xr3_band_SO}c, the main experimental features of the Pb-$\alpha$ phase DOS are reproduced by the calculations : one peak $O_1$ in the occupied-states, a depressed DOS $O_3$ around $E_F$ and a 3-peaks structure $U_2$, $U_3$, $U_4$ centered around the prominent peak $U_3$ in the empty-states. We underline that the spin-orbit splitting is about 50$\%$ of the Hubbard one demonstrating the crucial importance of including both effects in the bandstructure calculation.

To compare in more details theory and experiment, the Tersoff-Hamann (TH) simulation \cite{TersoffHamann} of the STS spectra is presented in Fig.~\ref{STM}e. A good qualitative agreement is found with the energies at which spectral features are observed in the occupied states. The agreement concerning their amplitude and shape is less good since $O_1$ is a kink in the experiment. Most likely this is because the TH approximation is not valid anymore close to  $O_1$ : Fig.~\ref{r3xr3_band_SO}c shows that states close to  $O_1$ have a large in-plane momentum $\|\bold{k}_{\parallel}\| \simeq \Gamma M'$. Thus  tip states with $l \neq 0$ angular momentum can significantly contribute to $dI/dV(V)$, invalidating the TH assumptions \cite{Chen2008}. Below $O_1$ in STS a broad peak with small oscillating features ends up around -0.43~eV in agreement with the calculated half-bandwidth. The small oscillations come from surface Si atoms (see Fig.~\ref{STM}g) and continue for $-0.70<V<-0.43$~eV (see Fig.~8 in \cite{supplemental}). The LDOS increases further for $V<-0.70$~eV due to bulk Si bands. For the unoccupied states, the main shape of a large peak $U_3$ with three additional features $U_1$, $U_2$ and $U_4$ agree with the calculations, although finite lifetime effects most likely broadens the real levels (see section VIII and Fig.~9 in \cite{supplemental}). Energies $U_1$, $U_2$, $U_3$, $U_4$ are found larger in STS due to band bending, as oberved in other isoelectronic systems \cite{Cortes,Modesti2007}. The measured upper half-bandwidth is 0.50~eV while the calculated one is 0.35~eV. One also cannot exclude that DFT+U slightly underestimates the bandwidth. Indeed HSE06 calculations show a larger bandwith, mostly for empty states (see \cite{supplemental}).

We now address the conductance oscillations measured between the Pb up and down atoms shown in Fig.~\ref{STM}f, which strongly support our interpretation of the LDOS structure. For electron energies $O_2<E<U_4$ DFT+U predicts a larger LDOS on the two Pb-down atoms than on the Pb-up ones (Fig.~\ref{STM}g). This situation is reversed for $E < O_2$. The local measurements presented in Fig.~\ref{STM}f and energy cuts in Fig.~\ref{STM}h agree well with these predictions in the range [-0.3;+0.4]eV. Finally, the experimental V-shaped DOS around $E_F$ on the meV scale is typical of correlated metals \cite{Lee1985} and can be modelled using dynamical Coulomb blockade \cite{Brun2012}.

As the Pb-$\alpha$ phase is a correlated 2D metal, it is interesting to probe whether it exhibits or not well-defined quasiparticle excitations at $E_F$. We answered this question by measuring quasiparticle interference (QPI) $dI/dV(V=0)$ map by STS and performing its Fourier transform (FT) \cite{SimonFTSTS}. In our case, we expect non-magnetic scalar scattering to be the most dominant scattering process. We verified that magnetic scatterers and spin-orbit scatterers do not contribute significantly to the FT-QPI maps using a similar approach as in Ref. \cite{Kohsaka}.

The calculated Fermi surface of the $3\times3$ reconstruction is shown in Fig. \ref{Fig:FTSTS}a. It consists of two hexagons, carrying essentially in-plane spin polarization with opposite helicities : it is thus almost of pure Rashba-type. We then calculated the joint DOS due to scattering by scalar impurities, namely $\chi(\bf q)=\frac{1}{N_K}\sum_{{\bf k}n,{\bf{k^{\prime}}}m} |M_{{\bf k}n, {\bf{k^{\prime}}}m}|^2 \delta({\epsilon_{{\bf k}n}-E_F) \delta(\epsilon_{{\bf k^{\prime}}m}}-E_F)$
where ${\bf q}={\bf k}-{\bf k^{\prime}}$,  $\epsilon_{{\bf k}n}$ are the relativistic DFT+U electronic bands and $M_{{\bf k}n, {\bf k^{\prime}}m}=\langle{\bf k}n|\sigma_0\exp\{i({\bf k}^{\prime}-{\bf k})\cdot {\bf r}\}|{\bf k^{\prime}}m\rangle$
with $\sigma_0$ the identity matrix in spin space. In the pure Rashba case one has
$M_{{\bf k}n, {\bf k^{\prime}}m}=1+(-1)^{n-m}\cos(\phi({\bf k})-\phi({\bf k}^{\prime}))$ and $\phi({\bf k})=(k_x+ik_y)/k$. In Fig. \ref{Fig:FTSTS}c we display $\chi(\bf q)$ as a colorplot. The hexagonal feature corresponds to interband $(m\ne n)$ scattering between segments of different hexagons of the Fermi surface, having parallel spin textures (see ${\bf q}_{\rm inter}$ in Fig. \ref{Fig:FTSTS}a).
Close to ${\bf q}={\bf 0}$ a star-shaped feature occurs due to intra Fermi sheet scattering along the straight side of each hexagon (see ${\bf q}_{\rm intra}$ in Fig. \ref{Fig:FTSTS}a). 

The 3-fold symmetrized experimental FT-QPI map is shown in Fig.~\ref{Fig:FTSTS}b (for raw data and details see IV in \cite{supplemental}). An hexagon of maximal radius $\simeq$~0.8~\AA$^{-1}$ is found in good agreement with Fig.~\ref{Fig:FTSTS}c. A smaller star-shaped feature near ${\bf q}={\bf 0}$ is also seen in the experiment. The good agreement between theory and FT-QPI maps demonstrates : (i) the existence of well-defined quasiparticles at $E_F$ forming two large hexagonal Fermi sheets with opposite in-plane spin-helicities, (ii) scattering mainly by scalar impurities. The remaining additional scattering signal probably comes from extended defects (see \cite{supplemental}).

In conclusion, by performing STS experiments and
first-principles calculations we have shown that the
Pb-substrate interaction, non-collinear spin-orbit coupling 
and correlation effects are all mandatory ingredients to correctly describe the electronic structure of the 1/3 ML Pb/Si(111) surface and its reconstructions. We have
found that the low-temperature ground-state of the $3\times 3$ reconstruction is a
strongly correlated metal with well-developed lower and upper Hubbard bands coexisting with a charge-density-wave phase. By comparing calculated quasiparticle interference with Fourier transform of STS  data, we demonstrated the occurrence of two large hexagonal Fermi surfaces
with in-plane spin polarization and opposite helicities. 


We acknowledge CINECA (ISCRA initiative), CINES, IDRIS and TGCC for computing resources. 
This work was supported by French state funds managed by the ANR within the Investissements d'Avenir programme under reference  ANR-11-IDEX-0004-02, and more specifically within the framework of the Cluster of Excellence MATISSE led by Sorbonne Universit\'es, RODESIS ANR-16-CE30-0011-01 and MISTRAL. We acknowledge M. Tringides, S. Pons, A. Tejeda, M. Cococcioni and A. Smogunov for useful discussions.

\bibliography{bibliography}{}

\end{document}